\documentclass[aps,pra,twocolumn,showpacs,floatfix]{revtex4}
\usepackage{epsfig}
\usepackage{graphicx}
\usepackage{dcolumn}
\usepackage{amsthm,amsmath}

\begin{document}
\title{Ba$^+$ Quadrupole Polarizabilities: Theory versus Experiment}
\author{B. K. Sahoo \footnote{Email: bijaya@prl.res.in}}
\affiliation{Theoretical Physics Division, Physical Research Laboratory, Ahmedabad-380009, India} 
\author{B. P. Das}
\affiliation{Theoretical Astrophysics Group, Indian Institute of Astrophysics, Bangalore-560034, India}
\date{Received date; Accepted date}
 
\begin{abstract}
Three different measurements have been reported for the ground state
quadrupole polarizability in the singly ionized barium (Ba$^+$) which 
disagree with each other. Our calculation
of this quantity using the relativistic coupled-cluster method disagrees 
with two of the experimental values
and is within the error bars of the other. We discuss the issues related to 
the accuracy of our calculations and emphasize the
need for further experiments to measure the quadrupole polarizability for this state and/or the $5D$ states.
\end{abstract}

\maketitle

\section{Introduction}
High precision studies of polarizabilities of atomic systems are of interest in a number of different problems
in physics \cite{itano,tatewaki,sherman,bks,quinet,banerjee, mahan,sternheimer}. A wide 
variety of methods have been used to calculate the polarizabilities of these systems \cite{mahan,sternheimer,bijaya1}.
Calculations of polarizabilities depend on the matrix elements between different atomic states
and the excitation energies between them. Therefore, the accuracies of these calculations
depend on the 
uncertainties in both the quantities. In general, it is very challenging to
minimize these uncertainties. However, in a sum-over-states approach \cite{arora,
bksra}, the major uncertainties in evaluating the polarizabilities can
be reduced by using the experimental energies. Furthermore, the accuracies
of the matrix elements can also be improved by matching the results of the lifetimes/branching ratios
of atomic states from sophisticated many-body calculations and high precision measurements.
For cases where the experimental results are not sufficiently accurate or if
all the available measured results are not in agreement, it is 
not possible to test the accuracies of the {\it ab initio} results. It might be
useful in such situations to employ the sum-over-states approach to evaluate 
the polarizabilities. 

For the ground state quadrupole polarizability in Ba$^+$, the available experimental
results \cite{gallagher,snow1,snow2} are neither in agreement
with each other nor with the calculations that are reported in this work.
Investigation of various properties using relativistic many-body methods 
for high precision studies in this ion are useful in the context of a proposed
parity nonconservation experiment \cite{fortson,bijaya2}, search for the nuclear
anapole moment \cite{bksanp}, estimation of the uncertainties for a proposed optical 
clock \cite{bkspra}, determination of the nuclear octuple moment \cite{lewty} etc.
In this paper, we report the results of our calculations of the matrix elements between different
atomic states of Ba$^+$ using the relativistic coupled-cluster (RCC) method. We also intend to
test the accuracies of some of the important matrix elements by using them to estimate the lifetimes of the $5D$ states and comparing with their corresponding measurements.

Before presenting our results, we define the quadrupole polarizability for a general atomic state 
in the following section. We give a brief description of the method of calculations of the wave
functions and the matrix elements in Section IV and then present our
calculated results and discussions after which we make our concluding remarks.

\section{Theory}
The potential energy of an atom in the presence of a static electric field is given 
by (for example, see \cite{mahan})
\begin{eqnarray}
V(r) &=&  - p_i \mathcal{E}_i - \frac{1}{6} Q_{ij} \partial_i E_j + \cdots \nonumber \\
  &=&  e \sum_i \mathcal{E}_i r^i P_i(cos \theta ),
\label{eqn1}
\end{eqnarray}
where $p_i$ and $Q_{ij}$ are the components of the electric dipole and quadrupole operators, 
respectively, $\mathcal{E}_i$ and $P_i(cos \theta )$ are the applied electric field
and Legendre polynomial, respectively, with component $i$. The quadrupole polarizability 
of an atomic state $\vert \Psi_n \rangle$ is related to the energy shift by the equation
\cite{dalgarno,maroulis}
\begin{eqnarray}
\delta E_n( \mathcal{E} ) =  - \frac{1}{8} \alpha^n_2 \mathcal{E}^2,
\label{eqn2}
\end{eqnarray}
where $\alpha^n_2$ is known as quadrupole polarizability of the state $\vert \Psi_n \rangle$ 
which is defined as
\begin{eqnarray}
\alpha^n_2 &=& - 2 \sum_{m \ne n} \frac{\left | \langle \Psi_n| {\rm{Q}} |\Psi_m\rangle \right|^2}{E_m - E_n}.
\label{eqn3}
\end{eqnarray}
and Q$=\sum q $ is the total electric quadrupole operator whose single particle reduced matrix element is
given by
\begin{eqnarray}
\langle \kappa_f\, ||\,q\,||\,\kappa_i \rangle &=& \langle \kappa_f\, ||\,C^{(2)}\,||\, \kappa_i \rangle 
\nonumber \\
&& \int_0^{\infty} dr \ r^2 \  (P_f(r)P_i(r)+Q_f(r)Q_i(r)), \ \ \ \ \ \
\end{eqnarray}
where $P(r)$ and $Q(r)$ represent the radial parts of the large and small components of the
single particle Dirac orbitals, respectively. The reduced Racah coefficients are given by
\begin{eqnarray}
\langle \kappa_f\, ||\, C^{(k)}\,||\, \kappa_i \rangle &=& (-1)^{j_f+1/2} \sqrt{(2j_f+1)(2j_i+1)} \ \ \ \ \ \
 \ \ \nonumber \\
                  &&        \left ( \begin{matrix}
                              j_f & k & j_i \cr
                              1/2 & 0 & -1/2 \cr
                                       \end{matrix}
                            \right ) \pi(l_{\kappa_f},k,l_{\kappa_i}), \ \ \ \ \
\end{eqnarray}
with
\begin{eqnarray}
  \pi(l,m,l') &=&
  \left\{\begin{array}{ll}
      \displaystyle
      1 & \mbox{for } l+m+l'= \mbox{even}
         \\ [2ex]
      \displaystyle        0 & \mbox{otherwise.}
    \end{array}\right.
\label{eqn4}
\end{eqnarray}

\section{Method of calculations}
In the present study, the atomic states of Ba$^+$ that we have considered have closed-shell 
cores and a valence electron $v$. We adopt a two step procedure to calculate
the wave functions for these states in the Fock space representation: First, the 
Dirac-Fock (DF) wave function for the common closed-shell core $[5p^6]$ (denoted by 
$|\Phi_0\rangle$) is calculated. In the next step, we append a corresponding
valence orbital $v$ to obtain the new DF wave function for the new configuration
(denoted by $\vert \Phi_v \rangle = a_v^{\dagger}|\Phi_0\rangle$). The atomic state 
for the new configuration is expressed using the RCC ansatz as
\begin{eqnarray}
|\Psi_v \rangle &=& e^T \{1+S_v\} |\Phi_v \rangle ,
\label{eqn5}
\end{eqnarray}
where $T$ and $S_v$ are the excitation operators that take into account correlation effects
arising from the core and valence electrons, respectively. We consider only the singly 
and doubly excited configurations from the
DF wave functions by approximating the $T$ and $S_v$ operators as
\begin{eqnarray}
T = T_1+T_2 \ \
\text{and} \ \ S_v = S_{1v}+S_{2v}.
\label{eqn6}
\end{eqnarray}
The above approximation is known as the coupled cluster singles and doubles (CCSD) method. Also, we construct triple excitation configurations from 
$\vert \Phi_v \rangle$ perturbatively in the spirit of the CCSD(T) approach
\cite{bijaya1,bijaya2,sahoo1,mukherjee,mukherjee1,lindgren,kaldor}.

The amplitudes for both the $T$ and $S_v$ operators are determined using the 
following equations
\begin{eqnarray}
\langle \Phi_0^K |\{\widehat{H_Ne^T}\}|\Phi_0 \rangle &=& \delta_{0,K} \Delta E_{corr}
\label{eqn7}
\end{eqnarray}
and
\begin{eqnarray}
\langle \Phi_v^K|\{\widehat{H_Ne^T}\} \{1+S_v\}|\Phi_v\rangle &=& \langle \Phi_v^K|1+S_v|\Phi_v\rangle 
\nonumber \\ && \langle \Phi_v|\{\widehat{H_Ne^T}\} \{1+S_v\} |\Phi_v\rangle \nonumber \\
 &=& [ \delta_{v,K} + \langle \Phi_v^K|S_v|\Phi_v\rangle ] \Delta E_v^{att}, \nonumber \\ 
\label{eqn8}
\end{eqnarray}
respectively. Here $K=1, \ 2 \cdots$ represents the singly, doubly etc excited configurations
with respect to their corresponding reference states, $\widehat{H_Ne^T}$ denotes the
connected terms of the normal order Dirac-Coulomb (DC) Hamiltonian ($H_N$) with the $T$ 
operators, $\Delta E_{corr}$ and $\Delta E_v^{att}$ are the correlation energy for the
closed-core and attachment energy of the valence electron $v$, respectively.

We evaluate the scalar polarizabilities by expressing them as the sum of three contributions
\begin{eqnarray}
\alpha^v_2  &=& \alpha^v_2(c) + \alpha^v_2(cv) + \alpha^v_2(v),
\label{eqn9}
\end{eqnarray}
where $\alpha^v_q(c)$ takes into account the contributions of the core orbitals, $\alpha^v_q(cv)$ 
and $\alpha^v_q(v)$ are the core-valence and valence contributions, respectively. In terms of the 
reduced matrix elements, the expressions of these parts are given by
\begin{eqnarray}
\alpha^v_2(c) &=& - \frac{2}{5} \sum_{c,p} \frac{\left | \langle J_p|| {\rm{Q}} || J_c\rangle \right|^2}{E_c - E_p}, \\
 \alpha^v_2(cv) &=& - \frac{2}{5 (2J_v+1)} \sum_{c} \frac{\left | \langle J_v|| {\rm{Q}} || J_c\rangle \right|^2}{E_c - E_v}, 
\end{eqnarray}
and
\begin{eqnarray}
\alpha^v_2(v) &=& - \frac{2}{5 (2J_v+1)} \sum_{m \ne v } \frac{\left | \langle J_v|| {\rm{Q}} || J_m\rangle \right|^2}{E_m - E_v} ,
\end{eqnarray}
where $\langle J_p|| {\rm{Q}} || J_c\rangle$ are the reduced matrix elements between the atomic
states with angular momenta $J_p$ and $J_c$. $\alpha^v_2(c)$ and $\alpha^v_2(cv)$ have 
been calculated using the third order many-body perturbation theory (MBPT(3) method) from 
an expression
\begin{eqnarray}
\alpha_2^n  &=&  \langle \Psi_n |{\rm{Q}} | \Psi_n^{(1)} \rangle,
\label{eq27}
\end{eqnarray}
where $\vert \Psi_n^{(1)} \rangle$ is like a first order perturbed wave function and it is 
obtained by solving the following inhomogeneous equation
\begin{eqnarray}
(H- E_n) \vert \Psi_n^{(1)} \rangle &=& (E_n^{(1)} - {\rm{Q}}) \vert \Psi_n \rangle,
\end{eqnarray}
with $E_n^{(1)} =  \left < \Psi_n |{\rm{Q}}|\Psi_n \right >$; which is similar to the first
 order perturbation equation. It should be noticed that unlike for the dipole operator
\cite{bijaya1,bpdas}, 
$E_n^{(1)}$ results are finite for the quadrupole operator and correspond to the quadrupole
moments of the respective states $|\Psi_n \rangle$.

Contributions from $\alpha^v_2(v)$ were determined by calculating important intermediate
states explicitly using the CCSD(T) method.
The reduced matrix elements between different states were computed using the following 
expression
\begin{eqnarray}
\langle J_f || {\rm{Q}} || J_i \rangle &=& \frac{\langle J_f || \{ 1+ S_f^{\dagger}\} \overline{{\rm{Q}}} \{ 1+ S_i \} || J_i\rangle}{\sqrt{{\cal N}_f {\cal N}_i}}, 
\end{eqnarray}
where $\overline{{\rm{Q}}}=e^{T^{\dagger}} {\rm{Q}} e^T$ and ${\cal N}_v = \langle \Phi_v |
e^{T^{\dagger}} e^T + S_v^{\dagger} e^{T^{\dagger}} e^T S_v |\Phi_v\rangle$ involve two 
non-truncating series in the above expression. The details of the calculations 
of these terms are discussed elsewhere \cite{bijaya1,bijaya2,sahoo1,mukherjee}.

\section{Results and discussions}
Below we present the quadrupole polarizabilities for the ground and the first two
excited $D$ states in Ba$^+$. We have used the experimental energies
from the National Institute of Science and Technology (NIST) data base \cite{nist} 
in the sum-over-states approach to evaluate the major contributions to the quadrupole polarizabilities, which
are the the valence correlation effects. The main purpose of doing
this is, as stated in the first section, to minimize the uncertainties in the calculated
results so that we shall be able to compare them meaningfully with the available experimental results.
Contributions from the higher excited states that cannot be accounted by the
valence correlation in the
the sum-over-states approach are evaluated using the MBPT(3) method; since
their contributions are typically smaller. The uncertainties in these results
are estimated by scaling results obtained using this method with the 
CCSD(T) calculations.

\begin{table}
\caption{\label{pol6s}The E2 matrix elements and the ground state quadrupole polarizability 
in Ba$^+$ in au. Results are given up to significant digits. Possible uncertainties in the 
results are given in the parentheses.}
\begin{ruledtabular}
\begin{tabular}{cccc}
\multicolumn{2}{c}{Transition} & \multicolumn{1}{c}{Amplitude} & \multicolumn{1}{c}{$\alpha^{E2}$} \\
\hline\\
\multicolumn{2}{c}{$\alpha_2^{6s}(v)$}  &       &  \\
$6s_{1/2} \rightarrow$ & $5d_{3/2}$ & 12.76(5) & 1466(11) \\
 & $6d_{3/2}$ & 16.58(12)  & 263(4)  \\
 & $7d_{3/2}$ &  5.727(7) & 24.07(6)  \\
 & $8d_{3/2}$ &  4.036(5) & 10.72(3) \\
 & $5d_{5/2}$ & 15.99(8) & 1978(20) \\
 & $6d_{5/2}$ & 19.99(20)  & 380(8) \\
 & $7d_{5/2}$ &  7.024(9) & 36.1(1) \\
 & $8d_{5/2}$ & 5.022(5)  & 16.59(3) \\
\multicolumn{2}{c}{$\alpha_2^{6s}(c)$} &    & 46(2)  \\
\multicolumn{2}{c}{$\alpha_2^{6s}(cv)$}  &  & $-0.001(0)$ \\
\multicolumn{2}{c}{$\alpha_2^{6s}({\rm{tail}})$}  & & $50(10)$ \\
 & & \\
\multicolumn{2}{c}{$\alpha_2^{6s}({\rm{total}})$} &  & 4270(27) \\
 & & \\
\multicolumn{2}{c}{Expt \cite{gallagher}} &  & 2050(100)\\
\multicolumn{2}{c}{Expt \cite{snow1}} &  & 2462(361) \\
\multicolumn{2}{c}{Expt \cite{snow2}} &  & 4420(250) \\
\multicolumn{2}{c}{Theo \cite{safronova}} &  & 4091.5 \\
\multicolumn{2}{c}{Theo \cite{safronova1}} &  & 4182(34) \\
\multicolumn{2}{c}{Theo \cite{patil}} &  & 4821 \\
 & & \\
\end{tabular}
\end{ruledtabular}
\end{table}
In Table \ref{pol6s}, we present the results of our quadrupole polarizability calculations of the ground state
of Ba$^+$. It is clear from this table that the dominant contributions (approximately 80\%)
come from the 5D states. Thus the accuracy
of the ground state quadrupole polarizability calculation depends primarily
on the accurate determination of the E2 matrix elements of the $6S_{1/2} \rightarrow 
5D_{3/2}$ and $6S_{1/2} \rightarrow 5D_{5/2}$ transitions. These matrix elements were
also calculated by us earlier using the same CCSD(T) method but with different basis
functions \cite{bijaya2,bijaya3}, and all the results are in good agreement.
There are also other calculations available for these matrix elements using different
variants of the RCC methods and basis functions \cite{geetha,sur,safronova,safronova1};
all the calculated results seem to be in reasonable agreement with each other. Moreover,
the accuracies of these matrix elements can be verified by using them to estimate
the lifetimes of the $5D$ states and comparing them with the measurements as
stated in the Introduction. Possible transition channels from the $5D_{3/2}$ state
to the ground state are due to the M1 and E2 multipoles. As we have shown in our earlier
work \cite{bijaya3}, the M1 transition probability is very small in this case and its
contribution to the lifetime of the $5D_{3/2}$ state is negligible. 
Therefore, neglecting the M1 contribution and using our calculated E2 matrix element 
of the $6S_{1/2}
\rightarrow 5D_{3/2}$ transition, we obtain the lifetime of the $5D_{3/2}$ state as
79.8(6) $s$. The measured values are 79.8(4.6) $s$ \cite{yu} and 89.4(15.6) $s$ 
\cite{gurell}. Our result is in good agreement with the first experimental result and is 
within the error bar of the second result. To calculate the lifetime of the 
$5D_{5/2}$ state, it is necessary to take into account all the M1 and E2 transition 
probabilities from this state to the $6S_{1/2}$ and $5D_{3/2}$ states. We have also
shown in Ref. \cite{bijaya3} that only the E2 and M1 transition probabilities
of the $5D_{5/2} \rightarrow 6S_{1/2}$ and $5D_{5/2} \rightarrow 5D_{3/2}$ 
transitions, respectively, are significant in the determination the lifetime of
the $5D_{5/2}$ state. In the present work, we have calculated the above M1 transition
amplitude to be 1.544(1) au which is in agreement with the result reported in 
Ref. \cite{bijaya3}. Using this value we find the lifetime of the $5D_{5/2}$ state to be 
29.8(3) $s$ with $84\%$ branching ratio to the $5D_{5/2} \rightarrow 6S_{1/2}$
transition. This is also in agreement with the experimental results which are
reported as 34.5(3.5) $s$ \cite{yu}, 31.6(4.6) $s$ \cite{gurell} and 32(2) $s$ 
\cite{plumelle}. We have used the experimental wavelengths
to determine the lifetimes of the $5D$ states in order to verify the accuracies
of the calculated E2 matrix elements. This analysis suggests that
our calculated E2 matrix elements are accurate to within a few percent and therefore
they can be considered for high precision studies of the quadrupole
polarizabilities of Ba$^+$.

\begin{table}
\caption{\label{pol5d3}The quadrupole polarizabilitiy of the $5D_{3/2}$ state of Ba$^+$ in au.
Uncertainties are given in the parentheses.}
\begin{ruledtabular}
\begin{tabular}{cccc}
\multicolumn{2}{c}{Transition} & \multicolumn{1}{c}{Amplitude} & \multicolumn{1}{c}{$\alpha^{E2}$} \\
\hline\\
\multicolumn{2}{c}{$\alpha_2^{5d3/2}(v)$}  &       &   \\
$5d_{3/2} \rightarrow$ & $6s_{1/2}$ & 12.76(5) & $-733(6)$  \\
& $7s_{1/2}$ & 4.882(5) & 13.96(3)  \\
& $8s_{1/2}$ & 1.558(2) & 1.0(2) \\
& $9s_{1/2}$ & 1.360(1) & 0.668(2) \\
& $6d_{3/2}$ & 8.36(4) & 37.3(5) \\
& $7d_{3/2}$ & 3.021(3) & 3.65(1) \\
& $8d_{3/2}$ & 2.183(2) & 1.69(1) \\
& $5d_{5/2}$ & 6.83(2) & 1278(7) \\
& $6d_{5/2}$ & 5.349(2) & 15.21(1) \\
& $7d_{5/2}$ & 1.978(2) & 1.56(1) \\
& $8d_{5/2}$ & 1.431(2) & 0.727(3) \\
& $5g_{7/2}$ & 8.45(8) & 26.9(5) \\
& $6g_{7/2}$ & 8.63(9) & 25.7(5) \\
\multicolumn{2}{c}{$\alpha_2^{5d3/2}(c)$}  &  & 46(2) \\
\multicolumn{2}{c}{$\alpha_2^{5d3/2}(cv)$} &  & $-0.49(3)$ \\
\multicolumn{2}{c}{$\alpha_2^{5d3/2}({\rm{tail}})$} &  & 116(30) \\
 & & \\
\multicolumn{2}{c}{$\alpha_2^{5d3/2}({\rm{total}})$} &  & 835(32) \\
 & & \\
\end{tabular}
\end{ruledtabular}
\end{table}

The next significant contributions to the ground state quadrupole polarizability 
come from the $6D$ states. It is difficult to estimate the accuracies of the E2
matrix elements of the $6S_{1/2} \rightarrow 6D_{3/2}$ and $6S_{1/2} \rightarrow 6D_{5/2}$ 
transitions from the measured lifetimes of the $6D$ states because of their negligible roles in the theoretical determination of these lifetimes.
Also, the contributions from the higher excited states (tail) and core correlations
to the final result of the quadrupole polarizability of the ground
state in Ba$^+$ are non negligible. A suitable method to test the validity of all these
contributions is to compare the final calculated result with the available measurements.
The reported experimental results for the ground state quadrupole polarizability of Ba$^+$ 
are 2050(100) \cite{gallagher}, 2462(361) \cite{snow1} and 4420(250) \cite{snow2} 
in au. The first two results agreewith  each other, but they are completely
in disagreement with the latest result. All the reported experimental results
have relatively large error bars. Two of these experimental results are just half of our calculated result. In such
a situation, it will not be possible to test the accuracies of the many-body methods that have been
used to perform the calculations as well as those that are
are likely to be developed in the {\it ab initio} framework to calculate these quantities. Also,
the above experimental techniques could single out the contributions from the 5D 
states to the ground state quadrupole polarizability and their values have been reported as
1562(93) \cite{shuman}, 2050(100) \cite{gallagher}, 1828(88) \cite{snow1} and 1524(8) 
\cite{snow2} in au. On the other hand using the E2 matrix elements of the $6S_{1/2} \rightarrow 5D_{3/2}$ and 
$6S_{1/2} \rightarrow 5D_{5/2}$ transitions, whose accuracies have been discussed above, 
we obtain the combined contributions of the $5D$ states to the ground state quadrupole
polarizability in Ba$^+$ as 3444(23) au with individual contributions as 1466(11) and 
1978(20) in au from the $5D_{3/2}$ and $5D_{5/2}$ states, respectively. It, therefore,
appears that the extracted values of these contributions of the $5D$ states to the
ground state quadrupole polarizabilities in the above experimental analysis may not 
be the contributions from both the $5D$ states, but rather from either the $5D_{3/2}$ or the
$5D_{5/2}$ state individually in the different experiments.

To the best of our knowledge, three calculations
of the ground state quadrupole polarizability in Ba$^+$ have been carried out, yielding 
4091.5 \cite{safronova}, 4182(34) \cite{safronova1} and 4821 \cite{patil} in au. All
these results are in agreement with our calculation; the first two are obtained
by the linearized RCC method (SDpT method), in contrast to our the non-linear approach, 
 but with the same level of hole-particle excitations from the DF wave functions. The two
results differ from each other by about 2\%, even though the same SDpT method was employed in the two cases,
and they are at the lower limit of the latest experimental result 4420(250) au. Their calculated results
are slightly lower than our result 4270(27) au and also, they do not 
overlap within their respective predicted uncertainties. The main reason for the
difference between the SDpT and our results is that the 
calculated values of the E2 matrix elements of the 
$6S_{1/2} \rightarrow 5D_{3/2}$ and $6S_{1/2} \rightarrow 5D_{5/2}$ transitions for the two cases are different
. It is necessary to emphasize that the precision of the theoretical determination of the lifetimes of the
$5D$ states depend critically on the accuracies of these matrix elements \cite{bijaya3, geetha, sur,safronova,safronova1}.
A precise measurement of the quadrupole polarizability of the ground state in Ba$^+$ could 
test the accuracies of these E2 matrix elements that are calculated by different many-body methods.
A third calculation of the 
ground state quadrupole polarizability has reported a result of 4821 au \cite{patil}
using rather simple wave functions based on the asymptotic behavior and the binding energy of the valence
electron in contrast to other calculations which are
based on all order perturbative methods in the RCC framework. Nevertheless, all the 
calculated results suggest that the ground state quadrupole polarizability in Ba$^+$
is approximately 4200 au and all the E2 matrix elements
of the $6S_{1/2} \rightarrow 5D_{3/2}$ and $6S_{1/2} \rightarrow 5D_{5/2}$ transitions
are in reasonable agreement with each other. In view of the discrepancies between the calculated and
experimental results as well as among the individual values of the latter, it would indeed be desirable to
perform precise measurements of the quadrupole
polarizability of the ground state of Ba$^+$.

\begin{table}
\caption{\label{pol5d5}The quadrupole polarizability of the $5D_{5/2}$ state in Ba$^+$ (in au). Estimated uncertainties to the results are mentioned in the parentheses.}
\begin{ruledtabular}
\begin{tabular}{cccc}
\multicolumn{2}{c}{Transition} & \multicolumn{1}{c}{Amplitude} & \multicolumn{1}{c}{$\alpha^{E2}$} \\
\hline\\
\multicolumn{2}{c}{$\alpha_2^{5d5/2}(v)$}  &       &   \\
$5d_{5/2} \rightarrow$ & $6s_{1/2}$ & 15.99(8) & $-659(6)$ \\
& $7s_{1/2}$ & 6.409(7) & 16.39(3)  \\
& $8s_{1/2}$ & 1.992(2) & 1.109(3) \\
& $9s_{1/2}$ & 1.723(2) & 0.724(2) \\
& $5d_{3/2}$ & 6.83(2) & $-852(5)$ \\
& $6d_{3/2}$ & 5.754(5) & 12.03(3) \\
& $7d_{3/2}$ & 2.030(2) & 1.114(2) \\
& $8d_{3/2}$ & 1.457(2) & 0.509(1) \\
& $6d_{5/2}$ & 11.26(2) & 45.8(2) \\
& $7d_{5/2}$ & 4.029(3) & 4.38(1) \\
& $8d_{5/2}$ & 2.925(3) & 2.05(1) \\
& $5g_{7/2}$ & 2.96(5) & 2.24(8) \\
& $6g_{7/2}$ & 3.00(6) & 2.1(1) \\
& $5g_{9/2}$ & 10.48(8) & 28.0(4) \\
& $6g_{9/2}$ & 10.62(9) & 26.3(5) \\
\multicolumn{2}{c}{$\alpha_2^{5d5/2}(c)$}  &  & 46(2) \\
\multicolumn{2}{c}{$\alpha_2^{5d5/2}(cv)$} &  & $-0.48(3)$ \\
\multicolumn{2}{c}{$\alpha_2^{5d5/2}({\rm{tail}})$} &  & 121(35) \\
 & & \\
\multicolumn{2}{c}{$\alpha_2^{5d5/2}({\rm{total}})$} &  & $-1201(36)$ \\
 & & \\
\end{tabular}
\end{ruledtabular}
\end{table}

An alternative approach to resolve this problem would be to measure the 
scalar quadrupole polarizabilities of the $5D$ states with high precision, and attempt to identify the individual
contribution of the the $6S$ state, perhaps in a manner 
similar to the method used in obtaining the contributions of the  the $5D$ states to the quadrupole polarizability
of the ground state. If that could be achieved, then it be possible to extract the E2 matrix elements of the 
$6S_{1/2} \rightarrow 5D_{3/2}$ and $6S_{1/2} \rightarrow 5D_{5/2}$ transitions. The advantages 
of measuring the scalar quadrupole polarizabilities of the
$5D$ states in addition to the ground state could be of two fold:
(i) the core-correlation effects, which is one of the significant contributions 
is the same for the states that we have considered, can be ignored in the estimation
of the accuracies of the E2 matrix elements and (ii)
By combining the quadrupole polarizabilities with the lifetimes of the
$5D$ states, it would be possible to extract the E2 matrix elements of the above two
transitions. For this purpose, we also calculate the quadrupole 
polarizabilities of the $5D$ states.

In Table \ref{pol5d3}, we present the scalar quadrupole polarizability of the $5D_{3/2}$
state. The largest contribution to this quantity comes from its
fine structure partner followed by the $6S$ state, but with opposite sign resulting in
a strong cancellation between them. The other significant contribution is from
the higher excited $g$ states (given as tail), which are not taken into account explicitly
in the sum-over-states approach. Also, we find the trend of the contributions to the
$5D_{5/2}$ state scalar quadrupole polarizability, given in Table \ref{pol5d5}, is 
similar to those for the $5D_{3/2}$ state but, the contributions from this state
and $6S$ state have same sign. The contributions from the high lying $g$ states are significant, and it would be appropriate to use a method that implicitly takes
all possible intermediate states into account, perhaps similar to Ref. 
\cite{mihaly}, for an accurate evaluation of the scalar quadrupole polarizabilities of the
$5D$ states. Nonetheless, the present study captures several classes of important 
correlation effects and it would be useful in guiding experiments on quadrupole 
polarizabilities in Ba$^+$.

\section{Conclusion}
We have carried out a detailed analysis of the calculated and experimental results of 
the reported ground state quadrupole polarizability in Ba$^+$ and highlighted 
the disagreement between the different studies. The
reported experimental results are not reliable enough to test the
validity of the calculated results. On the basis of different physical considerations, we propose
new theoretical and experimental studies of the quadrupole polarizabilities of the 5$D$ states to test the accuracies of the E2 matrix elements
between the ground state and the 5$D$ states and the quadrupole polarizability of the
ground state. We have also presented the results of our calculations for the quadrupole polarizability
of the 5$D$ states in Ba$^+$ using the E2 transition amplitudes obtained from our
calculations. 

\section*{Acknowledgement}
The calculations reported in this work were performed using the 3TFLOP HPC cluster computational facility at the  
Physical Research Laboratory, Ahmedabad.

\end{document}